\documentclass[pra, aps,eqsecnum,showpacs,amsmath]{revtex4}
\usepackage{amssymb}
\usepackage{amsfonts}
\usepackage{amsmath}
\newcommand{\epsdir}{./}

\providecommand{\fBOX}[1]{#1}

\newsavebox{\KAPTion}
\newtheorem{FIGURE}{{\rm Fig.}}
\providecommand{\FigInText}{}
\providecommand{\FigBelow}{}

\providecommand{\FigInText}[4]{}
\renewcommand{\FigInText}[4]{
\savebox{\KAPTion}{\parbox{#2mm}{
\begin{FIGURE}
\centering
\label{#3} {\rm #4 \fBOX{{\rm #3 }} } 
\end{FIGURE}}}
\begin{picture}(#2,5)(0,0)
\put(#1){\epsfxsize=#2mm\epsfbox{\epsdir#3.eps}}
\put(#1){\raisebox{-10mm}{\usebox{\KAPTion}}   }
\end{picture}
}

\providecommand{\FigBelow}[5]{}
\renewcommand{\FigBelow}[5]{
\savebox{\KAPTion}{\parbox{#2mm}{
\begin{FIGURE}
\centering
\label{#4} {\rm #5}
\end{FIGURE}}}
\begin{picture}(#2,#3)(0,0)
\put(#1){\epsfxsize=#2mm\epsfbox{\epsdir#4.eps}}
\put(#1){\raisebox{-10mm}{\usebox{\KAPTion}}  }
\end{picture}
}

\providecommand{\PutFig}[5][0]{}
\renewcommand{\PutFig}[5][0]{
\savebox{\KAPTion}{\parbox{#2mm}{
\begin{FIGURE}
\centering
\label{#4} {\rm #5}\  \ \fBOX{{\rm #4}}
\end{FIGURE}}}
\begin{picture}(#2,#1)(0,0)
\put(#3){\epsfxsize=#2mm\epsfbox{\epsdir#4.eps}}
\put(#3){\raisebox{-10mm}{\usebox{\KAPTion}}  }
\end{picture}
}

\input{epsf}
\begin{document}
\title{Hamilton's theory of turns revisited}
\author{N. Mukunda}
\email{nmukunda@cts.iisc.ernet.in}
\affiliation{Centre for High Energy Physics, Indian Institute of 
Science, Bangalore 560012}
\author{S. Chaturvedi}
\email{scsp@uohyd.ernet.in}
\affiliation{School of Physics, University of Hyderabad, 
Hyderabad 500046}
\author{R. Simon}
\email{simon@imsc.res.in}
\affiliation{The Institute of
 Mathematical Sciences, C. I. T. Campus, Chennai 600113}

\begin{abstract}
 We present a new approach to Hamilton's theory of turns for the groups 
 $SO(3)$ and $SU(2)$ which renders their properties, in particular their 
composition law, nearly trivial and immediately evident upon inspection. 
 We show that the entire construction can be  based on binary 
rotations rather than mirror reflections.
\end{abstract} 
\pacs{11.30.-j; 03.56.Ta; 03.65.Fd
{\bf Corresponding Author:} S. Chaturvedi~~ Phone: +91-40-2313 
4353; ~ Fax: +91-40-2301 0227}
\maketitle 

\section{Introduction}
Hamilton's theory of turns, arising out of  his profound and more 
general theory of quaternions, gives a beautiful  geometric 
visualisation of the elements of the groups $SO(3)$  and $SU(2)$, and of 
their {\em noncommutative} composition laws\,\cite{1}. Thus,  group 
elements 
can be pictured as {\em equivalence classes} of directed great  circle 
arcs on 
a two-dimensional sphere of turns, the equivalence being  with respect 
to sliding an arc along its great circle. The group composition  law is 
expressed by the tail-to-head `addition' of turns, reminiscent of  the 
{\em parallelogram law} of addition of free  vectors in the context of  
the abelian group of translations in an Euclidean space.

The important difference between the two situations is that while the 
propositions of Euclidean geometry are scale invariant, those pertaining 
to the sphere $\mathbb{S}^2$ are not. For instance, the unit sphere $\mathbb{S}^2$ does 
not support any 
notion of {\em similar triangles}; indeed, the angles of a spherical 
 triangle fully quantify also the area through the {\em spherical 
excess}. 
 In the present context these are hints  of the non-commutativity of 
the turns composition law. An important consequence of 
this lack of scale invariace is this: an $SO(3)$ rotation of amount 
$\alpha$ gets represented by a (geodesic) arc of length $\alpha/2$; 
no other multiple of $\alpha$ will return the correct multiplication law 
for the group in terms of geometric composition of these arcs.    

An easily  accessible and readable account of the theory of turns is 
given in the  well-known monograph of Biedenharn and Louck on angular 
momentum in  quantum physics\,\cite{2}. Applications to classical 
polarization  optics\,\cite{3}, to geometric phases for two-level 
systems\,\cite{4}, and generalisations to the noncompact groups 
$SL(2,\mathbb{R}) \sim SU(1,1) \sim Sp(2,\mathbb{R})$\,\cite{5} 
and $SL(2,\mathbb{C})$\,\cite{6} which are 
double covers of the Lorentz groups $SO(2,1)$ and $SO(3,1)$ 
respectively,   have been developed elsewhere. 

The purpose of this  paper is to present a treatment of turns which 
renders their origin  and properties extremely elementary, indeed to 
such an extent that  further simplification would seem essentially 
impossible. The main  property that is exploited is the fact that every 
plane rotation can be  expressed (not uniquely) as the product of two 
reflections, and 
the amount of freedom  available in doing so (this freedom or 
nonuniqueness turns out to be essential for the theory). It is 
then shown that the 
origin of the turn  concept can be traced to this geometrical fact, and 
that it can be easily  generalised to all proper rotations of 
$\mathbb{R}^3$. As a 
result 
of the formulae that  one is led to, the (noncommutative) rule for 
composing turns is found  to require no calculations at all as it is 
immediately evident upon inspection.

Section II  settles notations for describing elements of the two groups 
$SO(3)$ and  $SU(2)$ in a manner that matches the two--to--one 
homomorphism  from the latter to the former. Section III considers first 
the representation  of elements in the $SO(2)$ subgroup of $SO(3)$ as a 
product of two plane  reflections, and then generalises to all elements 
of $SO(3)$. As is known,  the concept of turns originates from such 
representations. However,  to render the group composition law as a 
trivial geometrical operation  with turns requires that we reexpress 
reflections in planes, which  are improper rotations, in terms of 
reflections through lines, which are {\em proper} rotations. These  are 
the so called binary rotations, 
and the details are given in  Section IV. The case of $SU(2)$ is taken 
up in Section V, and Section  VI contains some concluding remarks. 
 
\section{Notational preliminaries, the groups $SO(3)$ and $SU(2)$}

As is well known, $SO(3)$ and $SU(2)$ are locally isomorphic  compact 
connected three parameter Lie groups, which are moreover simple  at the 
level of their (common) Lie algebra. We describe the elements of  both 
groups using the well-known axis-angle parameters. Denoting axes  by 
unit vectors $\widehat{\bf{n}},\widehat{\bf{n}}^\prime,\cdots\in  
\mathbb{S}^2 $ and angles of (right handed) rotations by $\alpha,  
\beta,\cdots$, we can describe the (defining representation of the)  
real proper orthogonal rotation group $SO(3)$ in three Euclidean  
dimensions as follows:
\begin{eqnarray}
SO(3)&=&\{\,{\cal R}=3\times 3~{\rm real~matrix}\,|\,{\cal R}^T {\cal 
R}={\mathbb I}_{3\times 3},~{\rm det}{\cal R}=1\,\}\nonumber\\
&=&\{\,{\cal R}(\widehat{\bf{n}};\alpha)\,|\,\widehat{\bf{n}}\in 
\mathbb{S}^2,~\alpha\in [0,\pi]\,\},\nonumber\\
{\cal R}(\widehat{\bf{n}};\alpha)&=&\left(\,{\cal 
R}_{jk}(\widehat{\bf{n}};\alpha)\,\right),~~j,k=1,~2,~3,\nonumber\\
{\cal R}_{jk}(\widehat{\bf{n}};\alpha) 
&=&\delta_{jk}\cos\alpha+n_jn_k(1-\cos\alpha)
-\epsilon_{jkl}\,n_l\sin\alpha,\nonumber\\
&=&\left(\,e^{-i\alpha \widehat{\bf{n}}\cdot \bf{J}}\,\right)_{jk}, 
~~(J_l)_{jk}=-i\epsilon_{jkl}.
\label{2.1}
\end{eqnarray}
Group composition  is given by matrix multiplication.  While the angle 
$\alpha$ in ${\cal R}(\widehat{\bf{n}};\alpha)$ and
 ${\cal R}_{jk}(\widehat{\bf{n}};\alpha)$ can take any  real value, it 
is adequate to limit it to the range $[0,\pi]$ on account  of the easily 
verified relations
\begin{eqnarray}
{\cal R}(\widehat{\bf{n}};\alpha +2\pi) &=&{\cal 
R}(\widehat{\bf{n}};\alpha),\nonumber\\
{\cal R}(\widehat{\bf{n}};\alpha)&=&{\cal R} 
(-\widehat{\bf{n}};2\pi-\alpha),
\label{2.2}
\end{eqnarray}
provided  we allow all $\widehat{\bf{n}}\in \mathbb{S}^2$. In this way 
there is a  unique set of axis angle coordinates for each ${\cal R}\in 
SO(3)$, except  for elements belonging to a subset of measure zero. 
These are the  elements ${\cal R}\in SO(3)$ corresponding to right 
handed rotations  of amount $\pi$ about all possible axes 
$\widehat{\bf{n}}\in  S^2$, sometimes called binary rotations. The  
nonuniqueness of parameters  for such elements arises from the relations 
\begin{eqnarray}
{\cal R}_{jk}(\widehat{\bf{n}};\pi)&=&2n_jn_k-\delta_{jk},\nonumber\\
{\cal R}(\widehat{\bf{n}};\pi)&=& {\cal R}(-\widehat{\bf{n}};\pi),
\end{eqnarray}
a reflection of the nontrivial global (topological) structure of 
$SO(3)$. 
Clearly, binary rotations are square roots of the identity: 
\begin{equation}
{\cal R}(\widehat{\bf{n}};\pi){\cal R}(\widehat{\bf{n}};\pi)=\mathbb{I}_{3\times 3}.
\end{equation}
 The $SO(3)$ composition rule, i.e., the explicit determination of 
 $\widehat{\bf{n}}^{\prime\prime},\alpha^{\prime\prime}$ in terms of 
 $\widehat{\bf{n}},\alpha$ and 
$\widehat{\bf{n}}^{\prime},\alpha^{\prime}$ such that
\begin{equation}
 {\cal R}(\widehat{\bf{n}}^\prime;\alpha^\prime){\cal 
 R}(\widehat{\bf{n}};\alpha)={\cal 
 R}(\widehat{\bf{n}}^{\prime\prime};\alpha^{\prime\prime}),
\label{2.5}
\end{equation}
 was evidently first obtained by Rodrigues in 1840 using the geometry of 
spherical triangles on $S^2$\,\cite{7}. 

 The analogous definitions for the unitary unimodular group $SU(2)$ in 
two complex dimensions are: 
\begin{eqnarray}
SU(2)&=&\{\,{\cal U}=2\times 2 ~{\rm complex~matrix}\,|\,{\cal 
U}^\dagger 
{\cal U}=\mathbb{I}_{2\times 2},~{\rm det\cal{U}}=1\,\}\nonumber\\
&=&\{\,{\cal U}(\widehat{\bf{n}};\alpha)\,|\,\widehat{\bf{n}}\in 
\mathbb{S}^2,~\alpha\in [0,2\pi]\,\},\nonumber\\
{\cal U}(\widehat{\bf{n}};\alpha)&=&\,e^{-i\alpha 
\widehat{\bf{n}}\cdot 
\mbox{\boldmath$\sigma$}/2}\,= 
\cos\frac{\alpha}{2}-i\widehat{\bf{n}}\cdot 
\mbox{\boldmath   $\sigma$}\sin\frac{\alpha}{2}.
\end{eqnarray}
Here the $\mbox{\boldmath    $\sigma$}$'s are the standard triplet of 
Pauli matrices. The range of  the angle $\alpha$ is now $[0,2\pi]$ in 
contrast to the $SO(3)$ case,  on account of the following replacements
\begin{eqnarray}
{\cal U}(\widehat{\bf{n}}; \alpha +2\pi)&=&-{\cal 
U}(\widehat{\bf{n}};\alpha),\nonumber\\
{\cal U}(\widehat{\bf{n}}; \alpha+4\pi)&=&{\cal 
U}(\widehat{\bf{n}};\alpha),\nonumber\\
{\cal U}(\widehat{\bf{n}};2\pi)&=&-\mathbb{I}_{2\times 2},
\label{2.7}
\end{eqnarray}
for the previous eqs.\,$(\ref{2.2})$. The Rodrigues formulae mentioned 
in the case of eq.\,$(\ref{2.5})$ hold again, with suitable extensions, 
for the SU(2) composition law
\begin{equation}
\cal{U}(\widehat{\bf{n}}^\prime; 
\alpha^\prime)\cal{U}(\widehat{\bf{n}};\alpha)=\cal{U} 
(\widehat{\bf{n}}^{\prime\prime}
;\alpha^{\prime\prime}).
\end{equation}

$SU(2)$ is a two-fold cover of $SO(3)$.  The corresponding homomorphism 
$\phi$ `preserves' parameters in the sense  that, consistent with 
eqs.\,$(\ref{2.2})$ and $(\ref{2.7})$, we have: 
\begin{equation}
\phi :\;SU(2)\rightarrow SO(3): \;\phi({\cal 
U}(\widehat{\bf{n}};\alpha))={\cal R}(\widehat{\bf{n}};\alpha).
\label{2.9}
\end{equation}

\section{Rotations, reflections, and the origin of turns}

Rotations about the $z$-axis,  when $\widehat{\bf{n}}= 
\widehat{\bf{e}}_3$, form an SO(2) subgroup of $SO(3)$: 
\begin{eqnarray}
SO(2;\widehat{\bf{e}}_3)&=&\{\,{\cal R}(\widehat{\bf{e}}_3; 
\alpha)\,|\,\alpha\in [0,2\pi]\,\}~\subset SO(3),\nonumber\\
{\cal R}(\widehat{\bf{e}}_3;\alpha^\prime){\cal R} 
(\widehat{\bf{e}}_3;\alpha)&=&{\cal R}(\widehat{\bf{e}}_3; 
\alpha^{\prime\prime}),\nonumber\\
\alpha^{\prime\prime}&=&\alpha^\prime+\alpha~\;{\rm mod}\;2\pi.
\label{3.1}
\end{eqnarray}
Notice that since the axis $\widehat{\bf{e}}_3$ is kept  fixed, the 
range of $\alpha$ here is $[0,2\pi]$ and not $[0,\pi]$ as  
in eq.\,$(\ref{2.1})$. 
The action on the $x$ and $y$ coordinates in  the plane 
perpendicular to $\widehat{\bf{e}}_3$ is most compactly  expressed in 
complex form:
\begin{eqnarray}
\xi=x+iy&:&{\cal R}(\widehat{\bf{e}}_3,\alpha)\xi=\xi^\prime 
=e^{i\alpha}\xi,\nonumber\\
\left(\begin{array}{c}x^\prime\\y^\prime\end{array}\right) 
&=&\left(\begin{array}{cc}\cos\alpha&-\sin\alpha\\  
\sin\alpha&\cos\alpha\end{array}\right) 
\left(\begin{array}{c}x\\y\end{array}\right).
\end{eqnarray}
Let us now introduce the improper operation  ${\cal P}_0(\alpha)$ which 
is reflection within the $x$-$y$ plane about  the line passing through 
the origin $x=y=0$ and making an angle $\alpha$  in the positive sense 
with the $x$ axis as shown in Fig\,1. \\
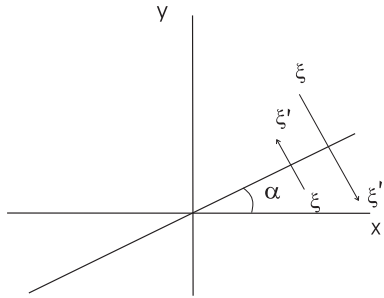
\FigBelow{50,10}{50}{50}{turnsfig1}{Showing the reflection ${\cal 
P}_0(\alpha)$}\\
In equations we have: 
\begin{eqnarray}
&&{\cal P}_0(\alpha)\xi=\xi^\prime=e^{2i\alpha}\xi^*,\nonumber\\
&&\left(\begin{array}{c}x^\prime\\y^\prime\end{array}\right) 
=\left(\begin{array}{cc}\cos2\alpha&\sin2\alpha\\ 
\sin2\alpha&-\cos2\alpha\end{array}\right) 
\left(\begin{array}{c}x\\y\end{array}\right).
\label{3.3}
\end{eqnarray}
The $2\times 2$ matrix ${\cal P}_0(\alpha)$ obeys 
\begin{eqnarray}
&&{\cal P}_0(\alpha)^T={\cal P}_0(\alpha)={\cal P}_0(\alpha+\pi),\nonumber\\
&&{\cal P}_0(\alpha)^2=\mathbb{I}_{2\times 2},\nonumber\\
&&{\rm det}{\cal P}_0(\alpha)=-1,
\label{3.4}
\end{eqnarray}
so we can limit $\alpha$ here to $[0,\pi)$.

The resultant of two such reflections about  two generally different 
lines is a proper rotation: 
\begin{eqnarray}
{\cal P}_0(\beta){\cal P}_0(\alpha)\xi  &=&{\cal 
P}_0(\beta) e^{2i\alpha}\xi^*\nonumber\\
&=&e^{2i(\beta-\alpha)}\xi,\nonumber\\
i.e., ~{\cal P}_0(\beta)
{\cal P}_0(\alpha)&=&{\cal R}(\widehat{\bf{e}}_3;2(\beta-\alpha)),\nonumber\\
i.e, {\cal R}(\widehat{\bf{e}}_3;\alpha)&=&{\cal P}_0 
(\beta+\alpha/2){\cal P}_0(\beta),{\rm any} \beta\in[0,2\pi). 
\label{3.5}
 \end{eqnarray}
We can represent this pictorially via the diagram in Fig\,2. It is 
clear that these reflections, for different values of $\alpha$, do not 
commute. It is for this reason that $O(2)$ is  non-abelian even 
though $SO(2)$ is an abelian group. \\
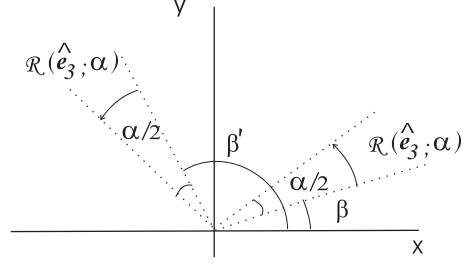
\FigBelow{50,15}{60}{50}{turnsfig2}{Showing the turn associated with  a 
rotation of the $x$--$y$ plane}\\
Remembering  that $\beta$ is arbitrary, we are immediately led to the 
representation  of the plane rotation  ${\cal 
R}(\widehat{\bf{e}}_3;\alpha)$ as a turn: a 
directed  (counter clockwise) arc of length $\alpha/2$ located anywhere 
on the unit  circle in the $x$-$y$ plane.
Again using  the `sliding' freedom -- equivalence relation among arcs -- 
 we recover the  composition law $(\ref{3.1})$ ,
\begin{eqnarray}
{\cal R}(\widehat{\bf{e}}_3;\alpha^\prime){\cal R}(\widehat{\bf{e}}_3; 
\alpha)&=&{\cal P}_0(\beta^\prime+\alpha^\prime/2){\cal 
P}_0(\beta^\prime) {\cal P}_0(\beta+\alpha/2){\cal P}_0(\beta),
~ {\rm any} \;\beta,\beta^\prime,\nonumber\\
&=&{\cal P}_0(\beta^\prime+\alpha^\prime/2) {\cal 
P}_0(\beta^\prime-\alpha/2),~~ \beta+\alpha/2=\beta^\prime, ~{\rm any} 
\;\beta^\prime,\nonumber\\
&=&{\cal R}(\widehat{\bf{e}}_3;\alpha+\alpha^\prime).
\end{eqnarray}
The  geometrical reflections construction of ${\cal 
R} (\widehat{\bf{e}}_3;\alpha)$ thus leads immediately  to the turns 
picture for such rotations. 

We will now generalize these considerations for rotations 
on a fixed plane to the full group of $SO(3)$ rotations, but it is 
useful to keep the following in mind. 
Since $SO(2)$ is abelian (and continuous), the double or indeed 
any multiple cover of $SO(2)$ remains isomorphic to $SO(2)$, displaying 
a kind of scale invariance. However when we pass on from $SO(2)$ to the 
full nonabelian group $SO(3)$, this feature is lost, just as the 
scale invariance of Euclidean space is absent on the unit sphere $\mathbb{S}^2$. 
   
The  extension to general ${\cal R}(\widehat{\bf{n}};\alpha)$ is 
 straightforward. For any mutually orthogonal vectors 
$\widehat{\bf{n}},\widehat{\bf{n}}_1\in \mathbb{S}^2$ we define 
\begin{eqnarray}
{\cal P}(\widehat{\bf{n}};\widehat{\bf{n}}_1)&=&
{\rm reflection,~in~plane~perpendicular ~to}~
\widehat{\bf{n}},\;{\rm 
in~line~}
\widehat{\bf{n}}_1\nonumber\\
{\rm i.e.},&& ~ {\rm reflection~about~the~plane~spanned~by~}  
\widehat{\bf{n}},\,\widehat{\bf{n}}_1.
\label{3.7}
\end{eqnarray}
Then as in eq.\,$(\ref{3.5})$ we easily obtain 
\begin{eqnarray}
{\cal R}(\widehat{\bf{n}};\alpha) &=&{\cal 
P}(\widehat{\bf{n}};\widehat{\bf{n}}_2) {\cal 
P}(\widehat{\bf{n}};\widehat{\bf{n}}_1),\nonumber\\
\widehat{\bf{n}}_1\cdot\widehat{\bf{n}}_2 
&=&\cos\frac{\alpha}{2},~~\widehat{\bf{n}}_1 
\wedge\widehat{\bf{n}}_2=\widehat{\bf{n}}\sin\frac{\alpha}{2}.
\label{3.8}
\end{eqnarray}
Clearly $\widehat{\bf{n}}_1,~ \widehat{\bf{n}}_2$  can be any two 
vectors perpendicular to $\widehat{\bf{n}}$, enclosing  angle 
$\alpha/2$, and such that $\widehat{\bf{n}}_1,  
~\widehat{\bf{n}}_2,~\widehat{\bf{n}}$ form a  right handed system. This 
gives the turns picture for the three-dimensional rotation ${\cal 
R}(\widehat{\bf{n}}; \alpha)$: 
\begin{eqnarray}
{\cal R}(\widehat{\bf{n}};\alpha)&=&{\rm any~directed~(by~ right-hand~ 
rule)~ arc~ of ~`length'} ~\alpha/2\nonumber\\ &&{\rm 
along~great~circle~on}~\mathbb{S}^2~{\rm perpendicular~to}~\widehat{\bf{n}}.
\end{eqnarray}
This is pictorially depicted in Fig\,3.\\
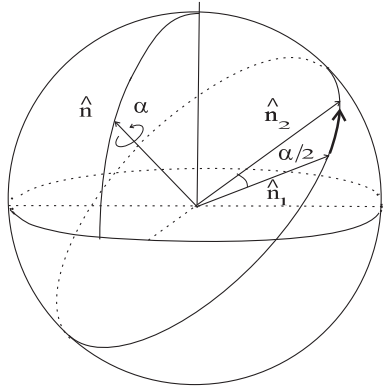
\FigBelow{60,10}{50}{65}{turnsfig3}{Showing the turn picture of a 
general rotation ${\cal R}(\widehat{\bf{n}};\alpha)$}\\

However this expression for a general ${\cal R}$  as a product of two 
${\cal P}$'s is not yet in a form convenient for  reading off a 
composition law for turns. For this we need to go  back from ${\cal 
P}$'s  to  ${\cal R}$'s.

\section{The composition law for turns}

Let us go back to the reflection  ${\cal P}_0(\alpha)$ of 
eqs.\,$(\ref{3.3}),(\ref{3.4})$ within the  $x$-$y$ plane. With the 
understanding that $z$ is invariant,  we can extend the $2\times 2$ 
matrix of ${\cal P}_0(\alpha)$ to a $3  \times 3$ matrix retaining the 
same symbol for simplicity: 
\begin{eqnarray}
{\cal P}_0(\alpha)&=&\left(\begin{array}{ccc}\cos2\alpha&\sin2\alpha&0\\\sin2\alpha&-\cos2\alpha&0\\0&0&1\end{array}\right),
\nonumber\\
{\cal P}_0(\alpha)^T&=&{\cal P}_0(\alpha),~~{\cal P}_0(\alpha)^2=\mathbb{I}_{3\times 3},\nonumber\\
{\rm det}{\cal P}_0(\alpha)&=&-1.
\end{eqnarray}
Therefore, since we are now in {\em odd  dimension}, we see that $-{\cal 
P}_0(\alpha)$ is a symmetric element of $SO(3)$, a proper binary rotation: 
\begin{equation}
-{\cal P}_0(\alpha)={\cal R}(\pm\sin\alpha,\mp\cos\alpha,0;\pi)
\end{equation}
both sign choices being allowed for binary rotations. For a general 
${\cal 
P}(\widehat{\bf{n}};\widehat{\bf{n}}_1)$ defined in  eq.\,$(\ref{3.7})$ 
we can then write the $3\times 3 $ matrix result
\begin{equation}
{\cal P}(\widehat{\bf{n}};\widehat{\bf{n}}_1) =-{\cal R}(\pm 
\widehat{\bf{n}}\wedge \widehat{\bf{n}}_1;\pi).
\end{equation}
( We emphasize that the geometric meaning of  the left hand side is that 
it is a reflection in three dimensions in the  plane perpendicular to 
$\widehat{\bf{n}}\wedge\widehat{\bf{n}}_1$, i.e.,  in the plane 
containing $\widehat{\bf{n}}$ and $\widehat{\bf{n}}_1$.  Thus points in 
this plane are unaffected. A binary rotation 
${\cal R}(\cdot,\pi)$ on the other hand is a reflection  in three 
dimensions through a line, not in a plane).

Putting this back into eq.\,$(\ref{3.8})$ we have:
\begin{eqnarray}
{\cal R}(\widehat{\bf{n}};\alpha)&=&{\cal P}(\widehat{\bf{n}};\widehat{\bf{n}}_2){\cal P}( \widehat{\bf{n}};\widehat{\bf{n}}_1)\nonumber\\
&=&{\cal R}(\pm \widehat{\bf{n}}\wedge \widehat{\bf{n}}_2;\pi){\cal R}(\pm \widehat{\bf{n}}\wedge \widehat{\bf{n}}_1;\pi),\nonumber\\
\widehat{\bf{n}}_1\cdot \widehat{\bf{n}}_2&=&\cos\alpha/2,~~\widehat{\bf{n}}_1\wedge \widehat{\bf{n}}_2=\widehat{\bf{n}}
\sin\alpha/2
\label{4.4}
\end{eqnarray}
Given  $ \widehat{\bf{n}}$ and $\alpha$ to begin with, all four choices 
of signs  are permitted, and $\widehat{\bf{n}}_1,\widehat{\bf{n}}_2$ can 
be chosen  freely subject to the conditions given. From the geometry 
involved,  we see that we can simplify this to say:
\begin{eqnarray}
{\cal R}(\widehat{\bf{n}};\alpha)&=&{\cal R}(\pm \widehat{\bf{n}}_2;\pi){\cal R}(\pm \widehat{\bf{n}}_1;\pi),~{\rm any~signs},\nonumber\\
\widehat{\bf{n}}_1\cdot \widehat{\bf{n}}_2&=&\cos\alpha/2,~~\widehat{\bf{n}}_1\wedge \widehat{\bf{n}}_2=\widehat{\bf{n}}
\sin\alpha/2
\label{4.5}
\end{eqnarray}
The turn  representing the element on the left runs from 
$\widehat{\bf{n}}_1 $towards $\widehat{\bf{n}}_2$. 

Now the  composition rule is immediate. Start with elements ${\cal 
R}(\widehat{\bf{n}};\alpha)$,\;${\cal 
R}(\widehat{\bf{n}}^\prime;\alpha^\prime)$  and assume 
$\widehat{\bf{n}}\neq\widehat{\bf{n}}^\prime$  for definiteness. In the 
sense of eq.\,$(\ref{4.5})$, let the pair $\widehat{\bf{n}}_1,  
\widehat{\bf{n}}_2$ go with the first element, and   
$\widehat{\bf{n}}_3, \widehat{\bf{n}}_4$ with the  second. Since the two 
great circles definitely intersect, we use the sliding  freedom to 
arrange $\widehat{\bf{n}}_2 =\widehat{\bf{n}}_3$ and then get:
\begin{eqnarray}
{\cal R}(\widehat{\bf{n}}^\prime;\alpha^\prime){\cal R}(\widehat{\bf{n}};\alpha)&=&{\cal R}(\widehat{\bf{n}}_4;\pi){\cal R}(\widehat{\bf{n}}_3;\pi){\cal R}(\widehat{\bf{n}}_2;\pi){\cal R}(\widehat{\bf{n}}_1;\pi)\nonumber\\
&=&{\cal R}(\widehat{\bf{n}}_4;\pi){\cal R}(\widehat{\bf{n}}_1;\pi),~\widehat{\bf{n}}_2=\widehat{\bf{n}}_3,\nonumber\\
&=&{\cal R}(\widehat{\bf{n}}^{\prime\prime};\alpha^{\prime\prime})
\end{eqnarray}
the `product' turn runs from $\widehat{\bf{n}}_1$ to  
$\widehat{\bf{n}}_4$.

\section{The SU(2) case}

We can develop a similar argument now, relying more on  algebraic 
relations than pure geometry. We begin with the general result 
\begin{equation}
({\bf a}\cdot\mbox{\boldmath{$\sigma$}})\,({\bf 
b}\cdot\mbox{\boldmath{$\sigma$}})= {\bf a}\cdot {\bf b}+i{\bf a}\wedge 
{\bf b}\cdot\mbox{\boldmath{$\sigma$}}
\end{equation}
involving the Pauli matrices. Now,  for a general element ${\cal 
U}(\widehat{\bf{n}};\alpha)\in SU(2)$  choose $\widehat{\bf{n}}_1$ and 
$\widehat{\bf{n}}_2$ as indicated in  eq.\,$(\ref{3.8}),(\ref{4.4})$, 
remembering however the extended range of $\alpha$. Then we find 
\begin{equation}
{\cal U}(\widehat{\bf{n}};\alpha)=-\,{\cal 
U}(\widehat{\bf{n}}_2;\pi)~{\cal U}(\widehat{\bf{n}}_1;\pi)
\label{5.2}
\end{equation}
which leads to the turns picture.  The elements ${\cal 
U}(\widehat{\bf{n}}^\prime;\pi)= 
-i\widehat{\bf{n}}^\prime\cdot\mbox{\boldmath{$\sigma$}}$  play the role 
of the earlier binary rotations, with the important difference  that 
their squares are $-\mathbb{I}_{2\times 2}$ as seen in the last  line of 
eqs.\,$(\ref{2.7})$. Now for the product rule:
\begin{eqnarray}
{\cal U}(\widehat{\bf{n}}^\prime;\alpha^\prime){\cal U} 
(\widehat{\bf{n}};\alpha)&=&{\cal U}(\widehat{\bf{n}}_4;\pi) {\cal 
U}(\widehat{\bf{n}}_3;\pi){\cal U}(\widehat{\bf{n}}_2;\pi){\cal U} 
(\widehat{\bf{n}}_1;\pi)\nonumber\\
&=&-\,{\cal U}(\widehat{\bf{n}}_4;\pi){\cal U}(\widehat{\bf{n}}_1;\pi), 
~\widehat{\bf{n}}_2=\widehat{\bf{n}}_3\nonumber\\
&=&{\cal U}(\widehat{\bf{n}}^{\prime\prime};\alpha^{\prime\prime})
\label{5.3}
\end{eqnarray}
The sliding  freedom has been used to arrange 
$\widehat{\bf{n}}_2=\widehat{\bf{n}}_3$,  and the turn for the product 
element then runs from $\widehat{\bf{n}}_1$  towards 
$\widehat{\bf{n}}_4$. As in the $SO(3)$ case,  the noncommutative 
`addition' of turns involves no calculations at all. 

\section{Concluding remarks}
The main step leading to trivialization of the  composition law for 
$SO(3)$ turns is the expression $(\ref{3.5})$ 
[\,equivalently $(\ref{4.5})$\,] of  a general element 
${\cal R}(\widehat{\bf{n}};\alpha)$ in factored form,  one factor each 
coming from the tail and the head of the turn.  (This also means that 
the great circle arc from $\widehat{\bf{n}}_1$  to $\widehat{\bf{n}}_2$ 
is drawn by us to help in visualization, as  eq.\,$(\ref{4.5})$  does 
not 
by itself require that it be drawn). The equation is also valid  as a 
relation between abstract group elements, i.e., it expresses a  property 
of the group in itself though we have obtained it through the  defining 
three dimensional representation. 

That every $SO(3)$ rotation is a  (nonunique) ordered pair of 
reflections about {\em  planes}  is known. That the same can be 
realized as an ordered  pair of binary rotations has the advantage that 
we stay within the $SO(3)$  group without having to make a `virtual 
transition' 
to the $O(3)$ group. A binary  rotation in three dimensions is 
reflection through an axis, 
but this reflection (unlike reflection in a plane) is an $SO(3)$ element.

In the case of $SU(2)$ relation $(\ref{5.2})$  holding in its defining 
representation, the representation at the level  of abstract group 
elements involves viewing the right hand side as  a product of three 
group elements; the negative sign on the right hand  side stands for 
$-\mathbb{I}_{2\times 2} $ in the defining representation,  and so for 
the nontrivial second element in the centre $ \mathbb{Z}_2$  of $SU(2)$ 
in the abstract. However the fact that this is in the centre,  and the 
property ${\cal U}(\widehat{\bf{n}};\pi)^2=
-\mathbb{I}_{2\times 2}$ mentioned after eq.\,$(\ref{5.2})$ ,  together 
ensure the proof 
of the turns composition law  $(\ref{5.3})$ 
 remains completely trivial. 

We may point out that binary rotations ${\cal R}(\widehat{\bf{n}};\pi)$  
in $SO(3)$ and the special elements ${\cal U}(\widehat{\bf{n}};\pi)$ in  
$SU(2)$ play similar roles in making composition law for turns trivial  
in each case. According to eq.\,$(\ref{2.9})$, the former are the  
results of the homomorphism $\phi:SU(2)\rightarrow SO(3)$ applied  to 
the latter.

The important difference between $SO(3)$ and $SU(2)$ needs mention.  In 
the case of $SO(3)$, every turn has `arc length' not exceeding $\pi/2$;  
while in $SU(2)$ we encounter turns of arc length upto $\pi$. When we  
use the turns composition rule for two  $SO(3)$ elements, $ {\cal 
R}(\widehat{\bf{n}};\alpha)$ and ${\cal R} 
(\widehat{\bf{n}}^\prime;\alpha^\prime)$, it  can happen that even 
though 
$\alpha/2,\alpha^\prime/2$ are both less than  or at most $\pi/2$, their 
`resultant' $\alpha^{\prime\prime}/2$ could exceed  $\pi/2$ as indicated 
in Fig\,4.\\
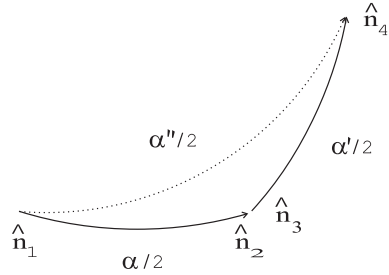
\FigBelow{50,10}{50}{50}{turnsfig4}{ Showing that the length of the 
composed turn can be as large as the sum of the lengths of the 
component turns}\vskip1.5cm

It is true that  $\alpha^{\prime\prime}/2$, being the length of the 
geodesic from $\widehat{\bf{n}}_1$ to 
 $\widehat{\bf{n}}_4$, is less than or equal to 
$\alpha/2+\alpha^{\prime}/2$, that is, 
 $\alpha^{\prime\prime}/2\leq \pi$; but this allows for 
$\alpha^{\prime\prime}/2 >  \pi/2$, i.e, we could have 
$\pi/2<\alpha^{\prime\prime}/2\leq \pi$.
However, in that case we can argue as follows:
 $$\frac{\alpha^{\prime\prime}}{2}>\frac{\pi}{2}\Rightarrow 
 \frac{(2\pi-\alpha^{\prime\prime})}{2}<\frac{\pi}{2}$$ 
and then 
 appealing to ${\cal R}(\widehat{\bf{n}};\alpha^{\prime\prime})={\cal 
 R}(-\widehat{\bf{n}};2\pi-\alpha^{\prime\prime})$ we can represent 
 the `product' turn also by an arc of length not more than $\pi/2$, but 
with  reversed sense. Such a problem is absent in the SU(2) case. 

This step which may be needed in the $SO(3)$ case motivates the 
following additional remarks. As is well known, $SU(2)$ and $SO(3)$ 
share a common Lie algebra as they are locally isomorphic. Nevertheless 
it is $SU(2)$ that is specially associated with this Lie algebra, in the 
sense that it is the unique simply connected Lie group arising from this 
Lie algebra. Globally as a manifold $SU(2)$ is the same as $\mathbb{S}^3$, while 
$SO(3)$ is $\mathbb{S}^3$ modulo the identification of `diametrically opposite' 
(or antipodal) points. (\,Thus $SU(2)$ is the double and universal 
covering group of $SO(3)$.\,) As a consequence, any (irreducible) matrix 
representation of the Lie algebra {\em always} exponentiates to an 
(irreducible) representation of $SU(2)$, which is faithful in only 
`half' the cases. It is only in the nonfaithful cases that we have an 
$SO(3)$ representation. It is these facts that ultimately underlie the 
comments made above in regard to turns for $SO(3)$. 

Given any two points $\widehat{\bf{n}}_1,\,\widehat{\bf{n}}_2\in \mathbb{S}^2$, 
we can always understand the angle between them, written as $\alpha/2$ 
as in Sections III, IV and V above, to be in the range $[0,\,\pi]$:    
 $\widehat{\bf{n}}_2 = \pm\widehat{\bf{n}}_1$ correspond to $\alpha/2= 
0,\,\pi$ respectively, otherwise $0<\alpha/2<\pi$. Now if we read 
eq.\,($\ref{4.5}$) from left to right, i.e., we start with some 
$SO(3)$ element  ${\cal R}(\widehat{\bf{n}};\alpha)$ where   
$\widehat{\bf{n}}\in S^2$, $\alpha\in [0,\,\pi]$ and find a pair   
$\widehat{\bf{n}}_1,\,\widehat{\bf{n}}_2$ for the right hand side, by 
our construction we will have $0\le\alpha/2\le \pi/2$. But if we read 
this equation `backwards' and start with a general pair of points    
$\widehat{\bf{n}}_1,\,\widehat{\bf{n}}_2$ on  $\mathbb{S}^2$: in case 
$0\le\alpha/2\le \pi/2$ we have the same situation as before; but in 
case $\pi/2<\alpha/2\le\pi$, we replace 
$\widehat{\bf{n}}_2$ by   
$\widehat{\bf{n}}_2^{\,\prime} = - \widehat{\bf{n}}_2$,   
${\cal R}(\widehat{\bf{n}}_2;\pi)$  by 
${\cal R}(\widehat{\bf{n}}_2^{\,\prime};\pi)$, and have the angle 
between   $\widehat{\bf{n}}_1$  and $\widehat{\bf{n}}_2^{\,\prime}$ 
back in the range $[0,\,\pi/2]$. This sign freedom is already explicitly 
stated in eq.\,($\ref{4.5}$). 

We can summarize by saying that turns are naturally or intrinsically 
associated with $SU(2)$, there being no restrictions in the choices of  
$\widehat{\bf{n}}_1, \widehat{\bf{n}}_2$ in eq. ($\ref{5.2}$); the angle 
$\alpha/2$ between them can be anywhere in $[0,\,\pi]$ as permitted by 
$S^2$.  Such restrictions show up only when we use turns to represent 
$SO(3)$ elements, so they must be carried along.

\end{document}